# Remote Sensing with High Spatial Resolution

**André Sandmann, Florian Azendorf, Michael H. Eiselt**
*Adtran Networks SE, Märzenquelle 1-3, 98617 Meiningen, Germany*
*andre.sandmann@adtran.com*

**Abstract:** Distributed fiber sensing based on correlation-aided phase-sensitive optical time domain reflectometry is presented. The focus is on correlation as an enabler for high spatial resolution. Results from different applications are presented. © 2024 The Author(s)

## 1. Introduction

The use of optical fibers to monitor their nearby environment is of high interest, since several Gigameters of fiber are already deployed in communication networks world-wide [1]. By separating the communication channels and sensing signals by wavelength, it allows for jointly performing both tasks at the same time. In addition, critical infrastructure such as water pipe networks, roads, electricity grids, gas and oil pipelines, as well as other infrastructure, are potential candidates for fiber-based sensing due to the fiber's resistance in harsh environments and potential monitoring distances beyond 100 km [2]. A technical summary of the latest work related to coherent optical time domain reflectometry (OTDR) aided by correlation is presented. The focus is on single-ended measurement of reflected and Rayleigh back-scattered signals.

In conventional OTDR systems, a single optical pulse is sent into the fiber. The longer the pulse duration, the more energy is transmitted into the fiber, resulting in a higher reach. However, the spatial resolution, i.e., the number of events that can be clearly distinguished within a 1-meter fiber, is inversely proportional to the duration of the probe pulse. By transmitting a code sequence into the sensor fiber and correlating the back-scattered received signal with the original transmitted code, the sequence is compressed into a narrow peak with the width of the bit duration, improving the spatial resolution while maintaining the same sensitivity compared to sending a long pulse. This enables the detection of closely spaced events at a large distance. In a direct-detection approach, correlation-aided OTDR can be used for a highly accurate propagation delay measurement with an accuracy of a few picoseconds, as demonstrated in [3]. With coherent detection, the phase information of the propagating optical signal can be extracted, which is highly sensitive to environmental changes. Considering a 1-meter optical fiber section to be strained by 1 micrometer, i.e., the application of 1 microstrain, leads to a propagation phase change of 4.7 rad. A high spatial resolution is necessary to avoid multiple $2\pi$-rotations, even for moderate strain values. Thus, combining correlation and coherent detection combines the advantages of high sensitivity to environmental effects and high spatial resolution at a high sensing reach.

An overview of coherent correlation OTDR (CC-OTDR) is provided in Section 2. In Section 3, experimental results in different application scenarios are presented, and Section 4 provides the concluding remarks.

## 2. Coherent correlation OTDR setup and challenges

In CC-OTDR, the optical power of a continuous wave (CW) laser is split into two parts. One part is modulated in a Mach-Zehnder modulator to generate the probing signal. We use a binary phase-shift keying modulated pulse sequence followed by a zero padding. Thus, to avoid ambiguities, only one pulse sequence propagates in the fiber at any time. After modulation, the probing signal is optically amplified and launched into the sensor fiber via an optical circulator. The probe signal is back-scattered along the sensor fiber, and the returned signal is transferred to a coherent receiver, where the second part of the CW laser is used as a local oscillator for homodyne detection. Amplitude and phase information are extracted for both received polarizations. After analog-to-digital conversion, the measured signals are cross-correlated with the original transmit sequence. The probe pattern therefore needs to be chosen properly to avoid large side lobes in the autocorrelation function, which might be misinterpreted as additional reflections.

As the probe signal and the local oscillator signal are emitted by the CW laser at different points in time, the coherent OTDR system requires an ultra-low phase noise laser, phase stable over the probe round-trip time, to be sure the measured phase variations stem from variations in the fiber and are not caused by laser phase noise. A further challenge is the handling of large data volumes, which requires suitable strategies for data reduction early in the processing chain. The evaluation of the measured results is often based on comparing the received signals from one probing shot to the signals from the previous shot. The interpretation of these changes and the corresponding classification of events can either be based on conventional models using physical considerations or can be aided by machine learning.

## 3. Application examples

A selection of three application examples interrogating different sensor fibers with the CC-OTDR is presented. Firstly, the focus is on the evaluation of a fiber Bragg grating (FBG) array with 2000 gratings spaced by 5 cm and having approximately the same center wavelength [4]. Figure 1 left shows the returned power versus position in a 2-m section of the fiber for different optical probing wavelengths, which are shown as offset to 1550 nm. The visible peaks are reflections from 39 gratings, and they show that each grating can be clearly separated. As a probe signal, we used two 2048-bit complementary Golay sequences at a bit rate of 5 Gbit/s, which provides a spatial resolution of 2 cm. By evaluating the peak power as a function of probe wavelength, the individual FBG reflection spectra can be measured, as illustrated in Fig. 1 right for 25 gratings. The results show a periodic variation in the Bragg wavelength for consecutive gratings. This variation has a periodicity of approximately 10 gratings. It has been found that these variations stem from periodic strain changes in the spooled fiber, and the 50-cm periodicity matches the circumference of the fiber spool. Unspooling the fiber significantly reduces the variations in the Bragg wavelength [4]. The presented approach involves tuning the laser wavelength, which takes several minutes for a full measurement and only allows for the measurement of slow strain and temperature variations of such fibers with an array of inscribed FBGs.

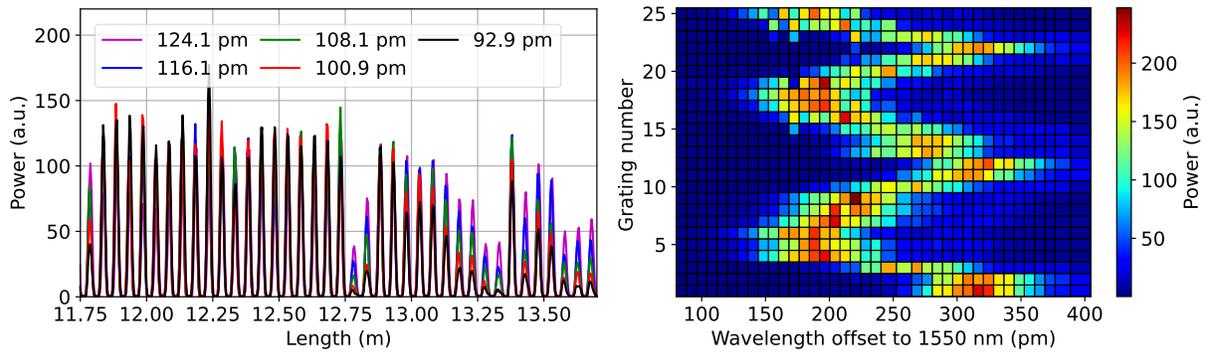

Fig. 1. FBG array evaluation with 2000 gratings and 5 cm spacing. left: Spatially resolved returned power of 39 gratings at different probing wavelengths − legend shows wavelength offset to 1550 nm; right: Reflection spectra of 25 gratings showing a periodic variation from spooling.

In the second application example, the effect of an acoustic signal applied to a standard single-mode fiber on the measured amplitude and phase information is demonstrated. A 418-meter fiber section is analyzed [5], for which Fig. 2a shows the averaged back-scattered power trace. The first peak at 8 meters originates from a connector reflection, and the peak at 418 meters is an open angled physical contact connector with a dust cap. In between the peaks, Rayleigh back-scattering is visible. The coherent superposition of signals back-scattered from various imperfections in the fiber leads to a distinct fading pattern of the Rayleigh back-scattering trace with minima and maxima, stemming from destructive or constructive interference, which is denoted as the fiber fingerprint. In the experiment, an acoustic signal of 120 Hz is applied to a 4-m section at approximately 216 meters from the fiber input connector. By evaluating the variation of the fiber fingerprint over time, as shown in Fig. 2b, variations in the region between 175 and 280 meters are visible, which show as interruptions of the vertical lines. However, it is difficult to exactly pinpoint the location of the applied acoustic signal and to identify the applied signal frequency based on this amplitude information. An analysis of the phase difference between neighboring back-scattering peaks is illustrated in a waterfall plot in Fig. 2c. Herein, the applied 120 Hz acoustic and vibration signal is clearly visible at 216 meters. The same measurement principle can be applied to longer fiber spans, as demonstrated in [6].

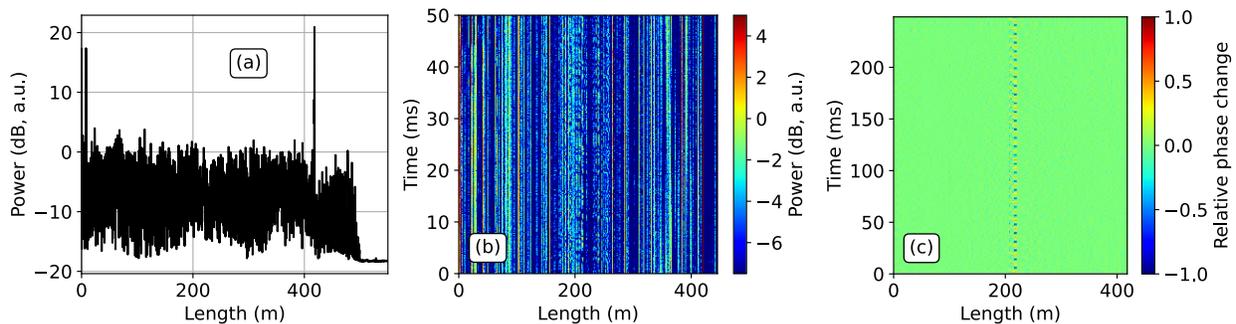

Fig. 2. Evaluation of a 418 meter sensor fiber with applying a 120 Hz acoustic signal. (a) Average fiber fingerprint; (b) Fingerprint change over time; (c) Phase waterfall plot showing the applied acoustic signal at 216 meters.

Figure 3 shows an example of measuring fast and slow effects that are simultaneously applied to a 195-meter fiber section [7]. Such measurements on different time-scales allow for the extraction of several types of environmental effects. Acoustics and vibrations are effects that occur on a short time-scale, in this example in the millisecond range. In contrast, temperature changes are comparably slow, e.g., in the order of seconds or minutes. In the experiment, the fiber section is placed inside a temperature chamber. Herein, a temperature profile starting at 30 °C, heating up to 40 °C, and cooling back down to 30 °C is applied. At the same time, a speaker in the same chamber provides a 400-Hz tone. When the temperature control of the chamber is active, a fan applies airflow to the fiber. The results in Fig. 3a show that, based on evaluating the phase difference, the applied 400 Hz tone can be clearly extracted. The applied heating leads to a quasi-linear phase slope on a short time-scale as shown in Fig. 3b. From the obtained phase slope and the reference temperature measurement, a refractive index variation with temperature of $10^{-5}$ 1/K is estimated. By analyzing the phase slopes, sampled at a rate of 2.7 seconds, the temperature on a long time-scale can be extracted, as shown in Fig. 3c. Herein, the black curve shows the reference temperature reading of the built-in sensor, and the blue line is the temperature measurement of the fiber core using the phase signal after calibration. Based on the core's temperature measurement, the chamber temperature is estimated by inverse filtering. The resulting red curve matches the reading of the built-in sensor. This example also highlights the high sensitivity of the optical phase to environmental effects.

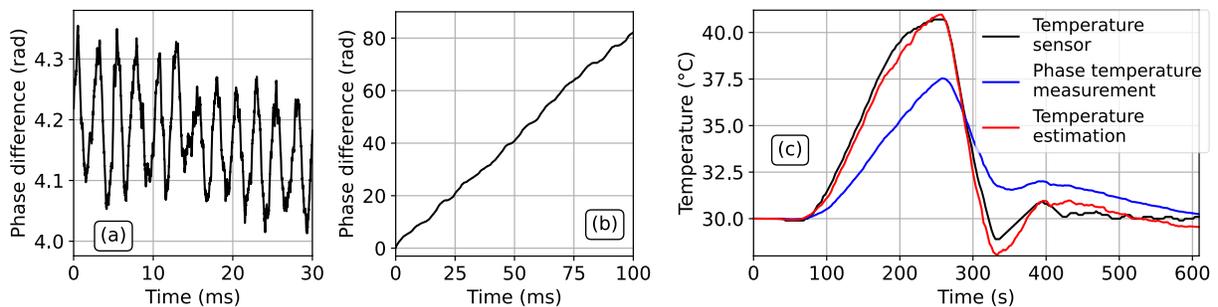

Fig. 3. Simultaneous acoustic and temperature measurements on different time-scales. (a) Phase difference signal with active speaker generating a 400 Hz tone; (b) Heating process with active fan and speaker; (c) Temperature measurement results.

## 4. Conclusion

An overview of remote fiber sensing with high spatial resolution is presented based on a coherent correlation OTDR. With correlation, the upper limit of the spatial resolution is determined by the symbol rate of the transmitted code sequence. Changing the symbol rate enables to flexibly change the configuration based on sensing needs. Three applications are presented where different effects on different time-scales are measured. These examples range from FBG array interrogation, showing static strain measurements with a 5 cm spacing, to the detection of dynamic strain from an acoustic signal applied to a standard single-mode fiber by evaluating changes in the optical phase. Finally, a simultaneous measurement of temperature and dynamic strain is shown.

**Acknowledgements**

This work has received funding from the Horizon Europe Framework Programme under grant agreement No 101093015 (SoFiN Project).